# Stability analysis of Bose Gas: effect of external trapping and three-body interaction with Gross-Pitaevskii equation in Tonks-Girardeau regimes

Rajmohan Sasireka[1], Olivier Tiokeng Lekeufack[2,*], Ambikapathy Uthayakumar[1], and Subramaniyan Sabari[3]

[1]Department of Physics, Presidency College (Autonomous), Chennai - 600005, India
[2]Department of Physics, Faculty of Science, University of Douala, P.O. Box 24157 Douala, Cameroon
[3]Instituto de Fisica Téorica, UNESP – Universida de Estadual Paulista, 01140-70 Sao Paulo, Brazil.

[*]Corresponding Author: lekeufackolivier@gmail.com

**Abstract:**

In this work, we investigate the stability aspects of quintic Gross-Pitaevskii (GP) equation with the presence and absence of external trapping potential for a Bose Gas (BG) in both the Tonks-Girardeau (TG) and the super Tonks-Girardeau (sTG) regimes. For this purpose, we compute both analytically and numerically a pure quintic GP equation with the presence of the scattering lengths. Using the time-dependent variational approach, we derive the equations of motion and effective potential of the system for both cases. Through the effective potential, we discuss the stability properties of pure quintic GP equation and obtain the modulational instability condition of BECs. The variational results are verified by means of direct numerical simulations using split-step Crank-Nicolson method and the observed results are in agreement with the analytical predictions.

**Keywords:** Stability analysis; Bose-Einstein Condensate; Gross-Pitaevskii equation; Tonks-Girardeau regimes.

**Statements and declarations:**
All authors certify that they have no affiliations with or involvement in any organization or entity with any financial interest or non-financial interest in the subject matter or materials discussed in this manuscript.

**Data availability Statements:**
All authors declare that there is no data to share.

I- Introduction

The quest for physics with ultracold atoms beyond the Bose-Einstein condensates (BECs) is one of the hottest topics in the area of atom optics today. After the first successful experimental realization of BECs in the year 1995 in rubidium [1], lithium [2], and sodium [3], a great deal of theoretical and experimental progress has been made in BECs and beyond the BECs limits. In the one-dimensional (1D) case, at strongly interacting regime, Bose gas (BG) exhibits a completely different behavior and BEC does not occur at finite temperature in a homogeneous system [4]. When the atomic density decreases, the interactions become increasingly dominant and the behavior of the system also changes. However, the wave functions of the atoms will become spatially distinct, and the bosons should start to repel each other, and essentially act like non-interacting fermions: this condition is known as the Tonks-Girardeau (TG) regime [5–7]. In the recently conducted experiments, it has been successfully achieved the TG regime and the resulting BG in this regime is so called TG gas [8,9]. The different physical regimes from the 1D BEC to 1D TG gas are usually characterized by the parameter $\gamma$, which is the ratio between interaction energy ($E_{int}$) and kinetic energy ($E_{kin}$). For high densities ($\gamma \ll 1$), the system is weakly interacting, a regime which can be described by mean-field theory, and in harmonically confined 1D systems Bose-Einstein condensation is possible. When the 1D density is decreased, the kinetic energy of the ground state is reduced and may get smaller than the interaction energy, thereby transforming the gas into a strongly interacting system, where the longitudinal motion of the particles is highly correlated. For ($\gamma \geq 1$) the correlation length becomes shorter than the mean interparticle distance. In the limit $\gamma \to \infty$, often referred to as the TG regime, a BG acquires fermionic properties due to the strong repulsive interactions. The different regimes associated with the parameter $\gamma$ are characterized by the excitation spectrum which can be probed by measuring the frequencies of collective oscillations [4,10–12].

The most recent experimental breakthroughs in this area are the realization of the stable highly excited gas-like phase called the super-TG (sTG) gas [13,14]. By using Feshbach resonances, suddenly changing from high repulsive interaction to high attractive interaction, the sTG gas was observed where the above properties were in agreement with the theoretical predictions [13,15]. Whereas the TG gas describes the strongly repulsive BG, the sTG gas describes a gas-like phase of the attractive BG which can be described by a system of attractive hard rods [15–19]. Now the strong repulsive and attractive interaction regime gases have been shown to be both theoretically exactly solvable and experimentally realizable [10,15–24].

Since the past decade, intensive investigations in strongly interaction regime have attracted a lot of attention from the theoretical as well as experimental point of view. Several studies have been carried out for describing the scaling law governing the expansion of the TG gas. Moreover, the 1D strong interaction regime gases have been proposed for the production of atomic clocks [20]. The strong contact interaction implies similarities between the bosonic system and a dual system of freely moving fermions, since the particle densities of the TG bosons and the dual system of fermions are actually equal. So, they may expect that the strong interaction regime is actually an asset for a good clock. Several groups have proposed different theoretical models for realizing the TG gases and analyzed the different properties of the TG gases using the methods like Bethe ansatz equation, Lieb-Liniger equation, Gaudin integral equation and hydrodynamic Luttinger equation [10–25]. One of the recent trends in this field is the application of mean-field-like approaches to describe the macroscopic dynamics of TG gases, a pure quintic Gross-Pitaevskii (QGP) equation with repulsive interaction [26]. Using the QGP equation, several works have been carried out in both the TG regimes by various groups [27–35]. In the present study, we have used QGP with both repulsive and attractive interatomic interactions for analyzing the stability of BG in the TG and sTG regimes respectively.

The effect of the interatomic interaction leads to a nonlinear term in the QGP equation while the s-wave scattering length, $a_s(t)$, plays an important role in the description of atom-atom interactions. The magnitude and sign of the s-wave scattering length, $a_s(t)$, can be tuned to any value, large or small, positive or negative by means of the "nonlinearity management"[36], i.e., the cubic nonlinearity periodically switching between self-attraction and repulsion, have been elaborated. This concept has been extended to the model containing the two- and three-body interactions [37,38], higher-order interactions [39-41], quantum fluctuations [42], dipolar interaction [43-45], and polariton [46] BEC regimes. The Feshbach resonances have been observed in gases of various alkali atoms. Currently, various research groups are using this technique to create a superfluid phase in degenerate Fermi gases also [47,48].

The organization of the present paper is as follows. In section II, we present a theoretical model equation for the present study. Then, we discuss the stability of BG in both TG regimes and point out the stability criteria for BG in the presence of external trap using variational method and split-step Crank-Nicolson (SSCN) method in section III A. In section III B, we study the stability criteria for BG in the absence of external trap using both analytical and numerical methods. Finally, we give the concluding remarks in section IV.

## II- NONLINEAR MODEL EQUATION

The original Gross-Pitaevskii equation (with two-body interaction) is widely used to describe a variety of properties of dilute Bose condensates for long-wavelength approximation, in particular trapped alkali gases [49]. But this approximation fails for short-wavelength cases. According to Kolomeisky and Straley, that quantitative features of Tonks (Fermi) density profile can be correctly addressed by the mean-field nonlinear QGP equation [26]. In the recent past, it has attracted a great deal of attention due to its potential applications in the theory of low dimensional condensed quantum gases. In the present study, we have used the following QGP equation to analyze the stability aspects of BG in both TG and sTG regimes [29–32]:

$$\left[-i\frac{\partial}{\partial t}-\frac{\partial^2}{\partial x^2}+V(x)\delta+\xi(t)|\psi(x,t)|^4\right]\psi(x,t)=0 \qquad (1)$$

Here, $\xi(t)$ = positive and negative values for TG and sTG regimes, respectively. $V(x)$ is an external harmonic potential and the parameter $\delta$ represents the strength of the external trap which can be switched to 1 or 0 for the BG in the presence of trap or in the absence of trap, respectively.

**III- STABILITY ANALYSIS**

**A. BG with external Harmonic trap**

In the present section, we study the stability properties of BG in both regimes with external trap. Recently, there have been great research interests with different trapping potentials [50]. For the mathematical simplicity and relevance to the recent experiments, we mainly focus on the impact of harmonic potential in the present section [50]. Taking into account the harmonic nature of the external trapping potential, we choose a Gaussian ansatz as an initial solution. Refs. [51,52] discussed the momentum distribution of noninteracting fermions and have shown that the fermionized bosons are equal for certain range of large interaction strength, while the localized wave packets of particle density bear a Gaussian form. In the following, we use the variational approach with the trial wave function (Gaussian ansatz) for the solution of equation (1) [49,53,54]:

$$\psi(x,t)=A(t)\exp\left[-\frac{x^2}{2R^2(t)}+\frac{i}{2}\beta(t)x^2+i\alpha(t)\right], \qquad (2)$$

where, $A(t)$, $R(t)$, $\beta(t)$ and $\alpha(t)$ are the amplitude, width, chirp and phase of the BEC, respectively. The Lagrangian density for equation (1) is given by

$$L=\frac{i}{2}\left(\frac{\partial\psi}{\partial t}\psi^*-\frac{\partial\psi^*}{\partial t}\psi\right)-\left|\frac{\partial\psi}{\partial x}\right|^2-\frac{x^2}{4}\delta|\psi|^2-\frac{1}{3}\xi|\psi|^6 \qquad (3)$$

The trial wave function equation (2) is substituted in the Lagrangian density and the effective Lagrangian is calculated by integrating the Lagrangian density as $L_{eff}=\int L\,dx$. We get:

$$L_{eff}=-\frac{3}{8\pi R^2(t)}-\frac{3R^2(t)}{32\pi}\delta-\frac{3R^2(t)\beta^2(t)}{8\pi}-\frac{\dot\alpha(t)}{4\pi}-\frac{3R^2(t)\dot\beta(t)}{16\pi}-\frac{\xi(t)}{36\sqrt{3}\pi^4 R^6(t)}, \qquad (4)$$

and the Euler-Lagrange equations for $R(t)$ and $\beta(t)$ are then obtained from the effective Lagrangian in a standard fashion as,

$$\dot R(t)=2R(t)\beta(t), \qquad (5)$$

$$\dot\beta(t)=-\frac{\delta}{2}+\frac{2}{R^4(t)}-2\beta^2(t)+\frac{4\xi(t)}{9\sqrt{3}\pi^3 R^8(t)} \qquad (6)$$

By combining equations (5) and (6), we get the following second-order differential equation for the evolution of the width,

$$\ddot{R}(t) = -R\delta + \frac{4}{R^3(t)} + \frac{8\xi(t)}{9\sqrt{3}\pi^3 R^7(t)}, \quad (7)$$

$$= -\frac{\partial U(R)}{\partial R}, \quad (8)$$

and the effective potential $U(R)$ corresponding to the above equation of motion can be written as

$$U(R) = \frac{R^2\delta}{2} + \frac{2}{R^2(t)} + \frac{4\xi(t)}{27\sqrt{3}\pi^3 R^6(t)}. \quad (9)$$

Now, we analyze the nature of the effective potential for BG in both TG and sTG regimes with repulsive and attractive interatomic interactions, respectively. Figure 1 depicts the potential energy curves as a function of R for different $\xi$ values by considering both TG and sTG regimes. In the presence of external trap case, the stability of the BG in the TG regime is achieved even when the strength of the interaction is very high in the system. Figure 1 (upper panel) illustrates the stability of BG in TG regime ($\xi$ = 50,100,150 and 250). But, from the lower panel of figure 1, the stability of the BG in sTG regime is lost while increasing the strength of the interaction in the system ($\xi$ =-10,-20,-30,-40 and -50). Next, the stability aspects of BG in both TG and sTG regimes have been analyzed numerically for time $t$ = 20 ms by directly solving the time-dependent QGP equation (1).

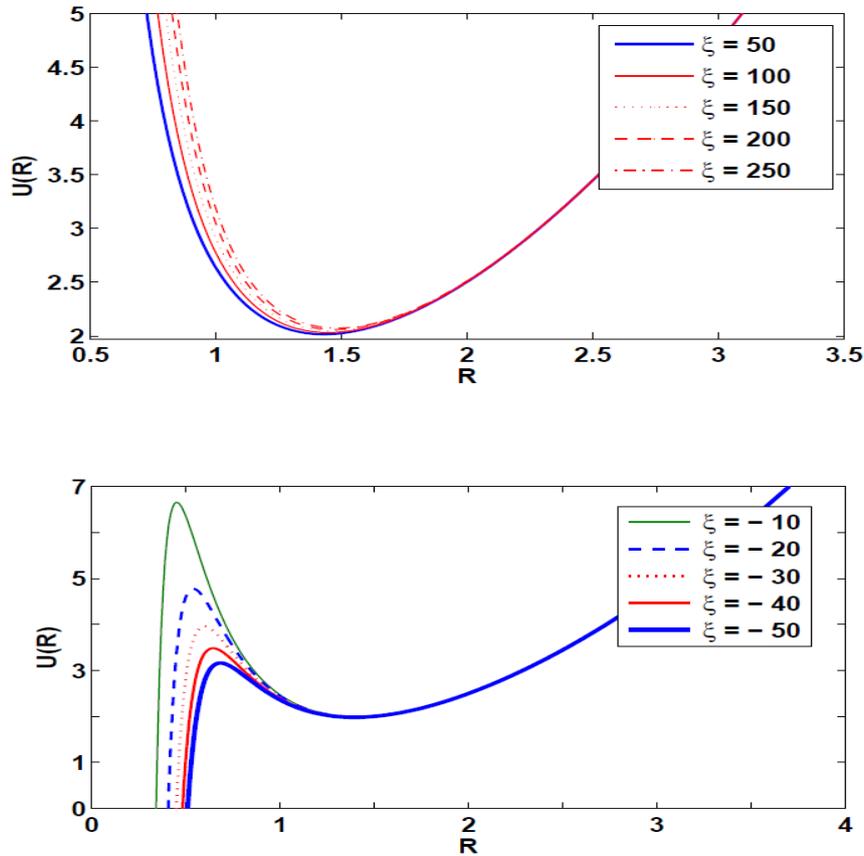

**Fig.1**: Plot of the effective potential $U(R)$ of equation (9) as a function of $R$ for $\delta$ (= 1) and different $\xi$ values. The upper (lower) figure shows the potential curve for TG (sTG) regime.

With the inclusion of external trapping, we set $\xi = 1$ in equation (1). The stability of the BG in TG regime is found to be maintained up to time $t = 20$ ms for $\xi = 50$. But in the case of sTG regime the stability of the BG is maintained only up to time $t = 1:5$ ms. This point is clearly illustrated in the spacetime plot and the rms radii of the BG in figure 2.

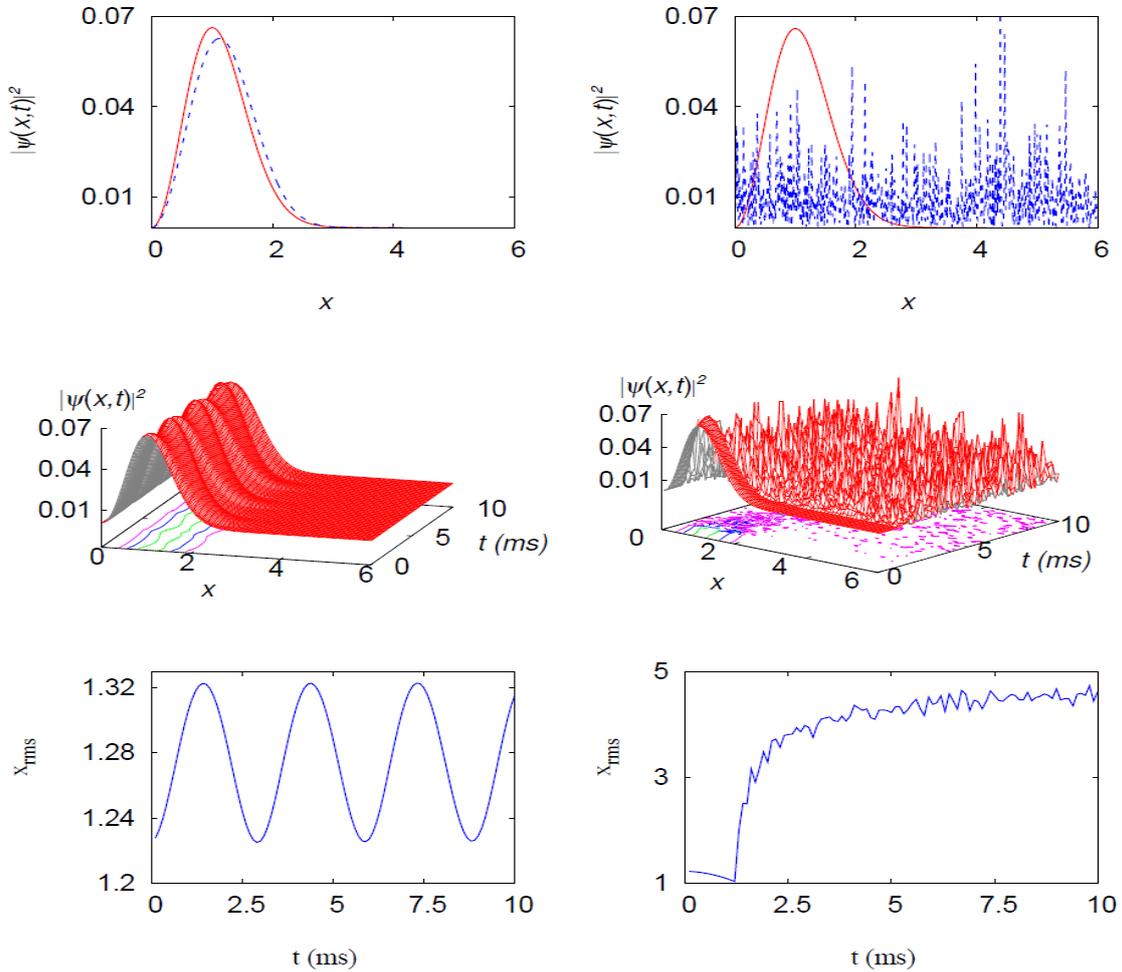

**Fig.2**: Numerical verification of figure 1. Left and right-side figures show the stability of the BG in TG ($\xi = 50$) and sTG ($\xi = -50$) regimes respectively, $\delta = 1.0$ in equation (1). The upper, middle and lower figures show the density profile, spacetime plot of the density and the rms radii of the BG respectively, which explains the stability of the BG in both TG and sTG regimes.

In figure 2, left side figures show the stability of the BG in TG regime. In this case the $x_{rms}$ of the condensate oscillates up to time $t = 20ms$, it shows that the BG in TG regime oscillates within the trap and the BG is stable up to time $t = 20ms$. The spacetime plot of the density $|\psi(x,t)|^2$ explains the oscillations of the BG in TG regime. But, in the sTG regime, the BG collapses for same magnitude ($\xi = -50$). As for the right-side plots in figure 2, the stability of the BG in the sTG regime is presented. In the sTG regime, the BG collapses due to the high interatomic attraction. In general, in the presence of trap, the stability of BG in TG and sTG regimes mainly depends on two important forces, one is the interatomic interaction and another one is the attraction due to the external trapping potential. In the TG regime, the attraction due to the external trap is compensated by the repulsive interatomic interaction between atoms in the TG gas and the stability of the TG gas remains stable. One can stabilize the BG in sTG regime by reducing the strength of the interatomic interaction or the strength of the external trap. So, we mainly concentrate our studies on the stabilization of BG in the sTG regime. In the case of sTG gas the system is stable only for very small strength of the interaction. When the interaction is considerably large, the system collapses due to the high interatomic attraction. If we reduce the strength of the attractive interaction then the stability of the system is increased. In the present study, we have shown that the stability of the sTG gas is increased by tuning the strength of the external trapping potential alone.

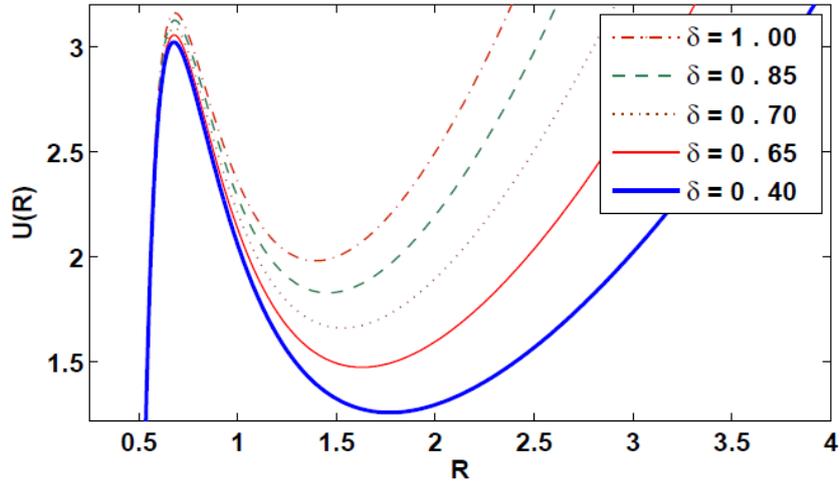

**Fig.3:** Plot of the effective potential $U(R)$ of equation (9) as a function of $R$ for BG in sTG regime ($\xi$ = -50) and different trapping strength ($\delta$) values.

Figure 3 illustrates the potential curve for various strength of the external trapping potential $\delta$ values. By decreasing the trapping strength for fixed number of atoms in the condensate or fixed strength of the three-body interaction ($\xi$) in the condensate, the depth of the minimum in the effective potential increases. It seems that the stability of the sTG gas is increased due to a decrease of the strength of the external trap. Figure 3 also illustrates the stability of the BG in both TG and sTG regimes for $|\xi|$ = 50. In the present study, we can stabilize the BG in the sTG regime without changing the interatomic force (or without reducing the number of atoms from the condensate). This can also be achieved by tuning the strength of the external trap. Figure 4 shows the stability of the BG in sTG regime for different trapping strengths.

To understand the stability of the BG by means of the rms size of the radius $X_{rms}$, we plot in Figure 5 the mean-square sizes $X_{rms}$ vs $t(ms)$ for different $\delta$ values. The oscillation of the $X_{rms}$ radii is related to the radial oscillation of the BG. In both cases, we study the resultant oscillation of the $X_{rms}$ radius of the BG. As clearly seen from figure 3 and also observed from figures 4 and 5, the stability of condensates increases by decreasing the strength of the external trap. Although the peak density oscillates with respect to time within the external trap, the density remains stable without breaking. Hence, the splitting of the density profile is represented as collapse of the condensation. It is clearly shown in figure 4 and figure 5 that the analytical solution of figure 3 is verified through numerical simulation. Hence, we conclude that one can stabilize sTG gas in the presence of external harmonic trap and the stability of the sTG gas can be increased by tuning the strength of the external trapping potential.

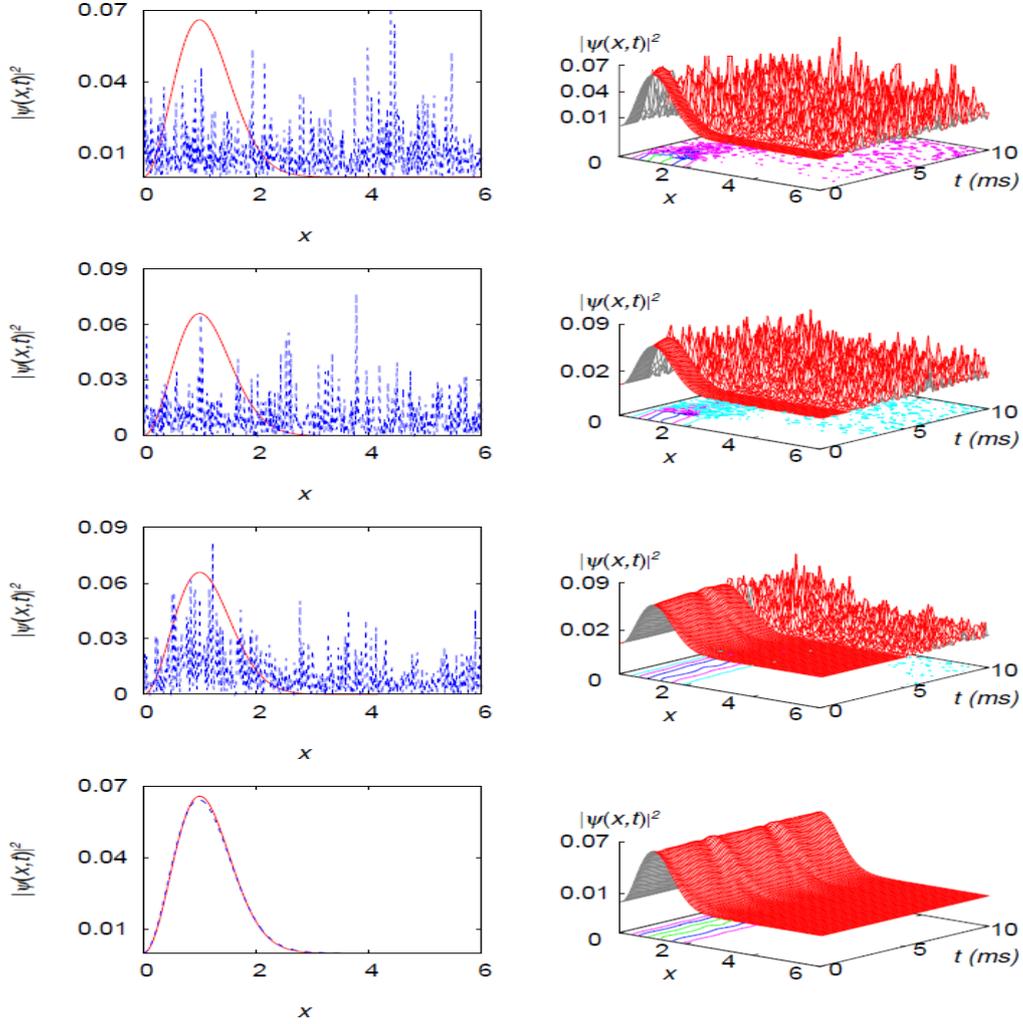

**Fig.4:** Numerical verification of figure 3. Left and right-side figures show the density profile and spacetime plot of the density with $\xi = -50$ in equation (1). Upper to lower figures for $\delta = 1.0$, $\delta = 0.9$, $\delta = 0.8$ and $\delta = 0.799$ respectively.

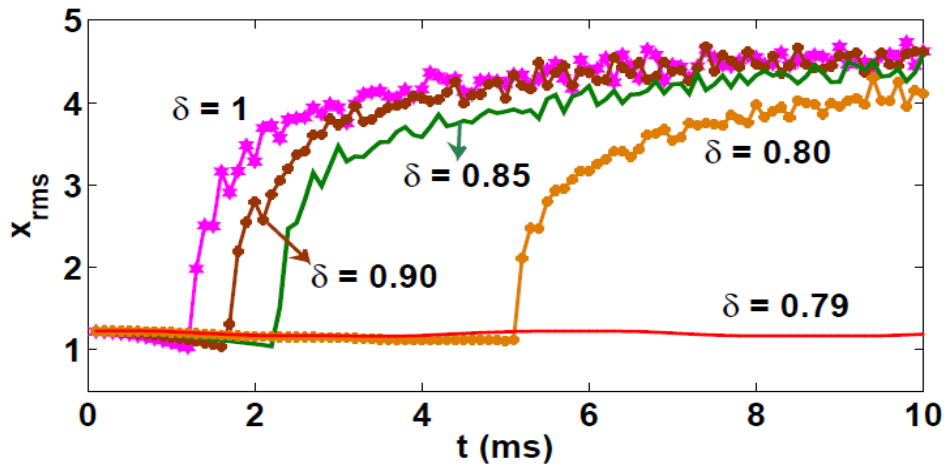

**Fig.5:** Numerical plot of the rms radii versus $t(ms)$ for $\xi = -50$ and different $\delta$ values in equation (1).

**B. Trapless BG**

Now, we analyze the stability of trapless BG in both regimes. Few research papers have already reported the trapless study using GP equation with many components and in several BEC regimes [36,37,39,41-46].

Hence, in the present section, we are interested to analyze the stability of trapless BG in TG regimes. When $V(x)=0$, the Lagrangian density for equation (1) is reduced to

$$L = \frac{i}{2}\left(\frac{\partial \psi}{\partial t}\psi^* - \frac{\partial \psi^*}{\partial t}\psi\right) - \left|\frac{\partial \psi}{\partial x}\right|^2 - \frac{1}{3}\xi|\psi|^6 \qquad (10)$$

The trial wave function equation (2) is substituted in the Lagrangian density (equation 10) and we get the following second-order differential equation for the evolution of the width:

$$\ddot{R}(t) = \frac{4}{R^3(t)} + \frac{8(\xi_0 + \xi_1 \sin(\omega t))}{9\sqrt{3}\pi^3 R^7(t)}, \qquad (11)$$

with $\xi(t) = \xi_0 + \xi_1 \sin(\omega t)$, where $\xi_0$ and $\xi_1$ are the amplitudes of constant and oscillating parts of the scattering length. Now, $R(t)$ can be separated into a slowly varying part $A(t)$ and a rapidly varying part $B(t)$ by choosing $R(t) = A(t) + B(t)$. When $\omega \gg 1$, $B(t)$ becomes of the order of $\omega^{-2}$. Keeping the terms of the order of $\omega$ up to $\omega^{-2}$ in $B(t)$, one may obtain the following equations of motion for $A(t)$ and $B(t)$ [38,55],

$$\ddot{B}(t) = \frac{8\xi_1 \sin(\omega t)}{9\sqrt{3}\pi^3 A^7(t)}, \qquad (12)$$

$$\ddot{A}(t) = \frac{4}{A^3(t)} + \frac{8\xi_0}{9\sqrt{3}\pi^3 A^7(t)} - \frac{56\xi_1 F(t)}{9\sqrt{3}\pi^3 A^8(t)}, \qquad (13)$$

where $F(t) = -\langle B(t)\sin(\omega t)\rangle$ indicates the time average of the rapid oscillation. From equation (12), we can get $B(t) = -\frac{8\xi_1 \sin(\omega t)}{9\sqrt{3}\pi^3 \omega^2 A^7(t)}$, and substituting it into equation (13), we obtain the following equation of motion for the slowly varying part,

$$\ddot{A}(t) = \frac{4}{A^3(t)} + \frac{8\xi_0}{9\sqrt{3}\pi^3 A^7(t)} + \frac{224\xi_1^2}{243\pi^6 \omega^2 A^{15}(t)}, \qquad (14)$$

$$= -\frac{\partial U(A)}{\partial A}, \qquad (15)$$

and the effective potential $U(A)$ corresponding to the above equation of motion can be written as,

$$U(A) = \frac{2}{A^2(t)} + \frac{4\xi_0}{27\sqrt{3}\pi^3 A^6(t)} + \frac{16\xi_1^2}{243\pi^6 \omega^2 A^{14}(t)}, \qquad (16)$$

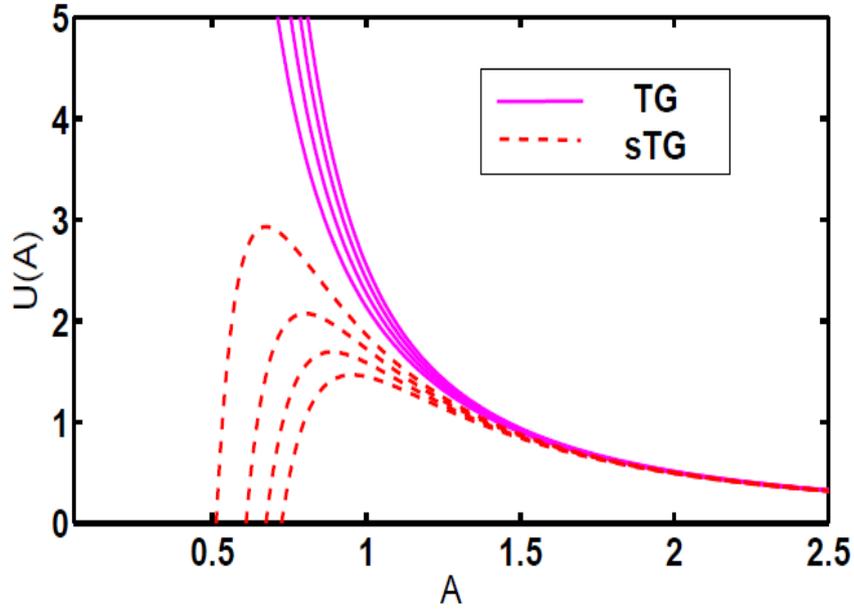

**Fig.6**: Plot of the effective potential $U(A)$ of equation (16) as a function of $A$ for $\xi_1 = 0$ in equation (16), where, (continues lines) $\xi_0 = 50, 100, 150, 200$ for TG gas and (dashed lines) $\xi_0 = -50, -100, -150, -200$ for sTG

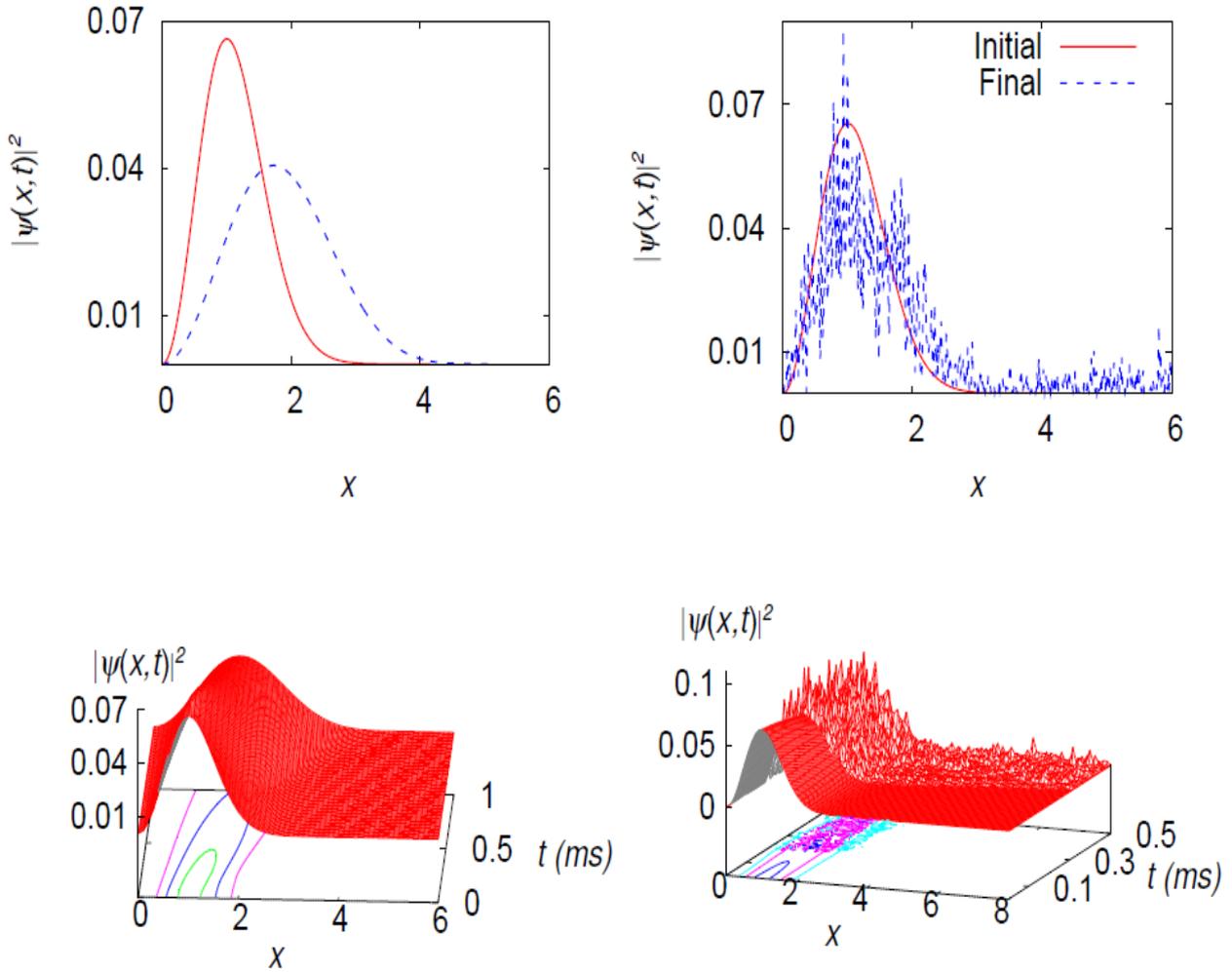

**Fig.7:** Stability of trapless BG in both TG and sTG regimes with $\delta = 0$ and $\omega = 0\pi$ in Equation (1), (left side) $\xi_f = 200$ for TG regime and (right side) $\xi_f = -200$ for sTG regime.

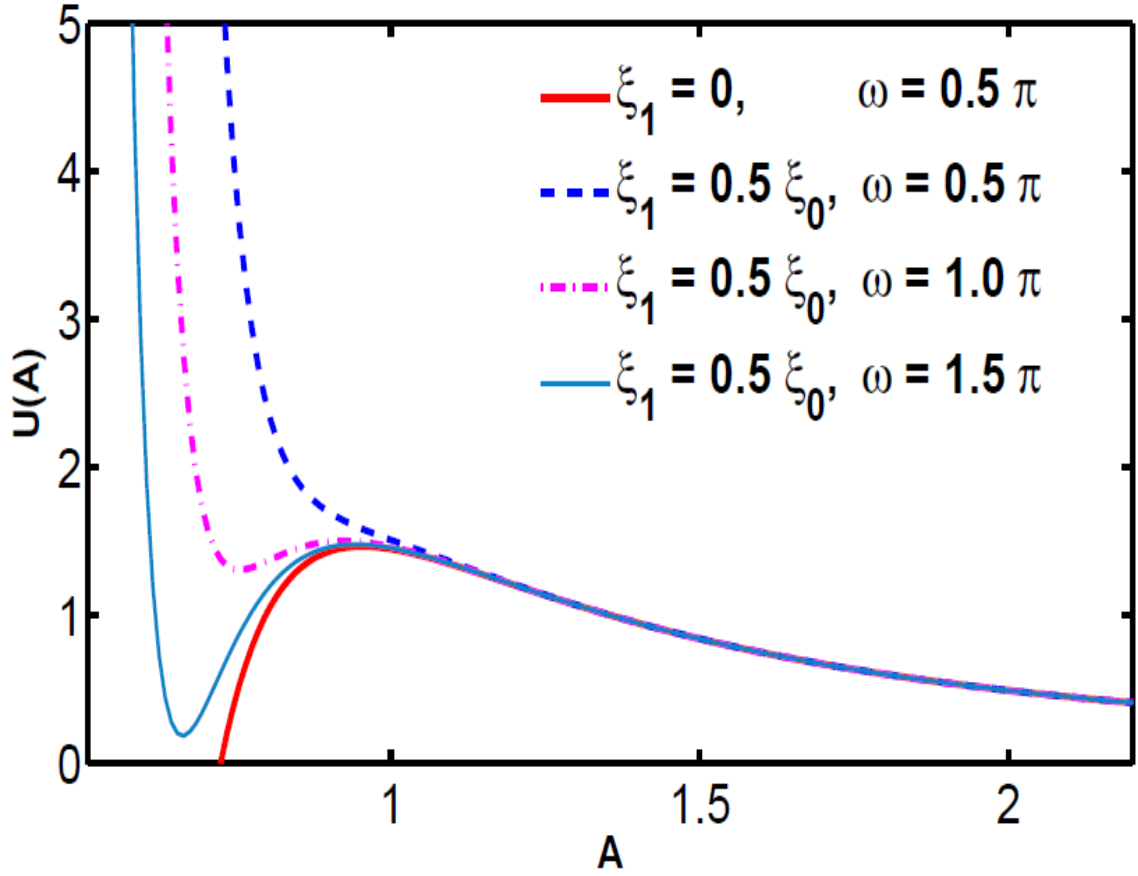

**Fig.8**: Plot of the effective potential $U(A)$ of equation (16) as a function of $A$.

Now, we analyze the nature of the effective potential for various strength of interactions, especially in the presence and absence of constant and oscillatory parts of interaction. Figure 6 depicts the potential energy curves as a function of the distance $A$ for the absence of oscillation part ($\xi_1 = 0$ in equation (16)). The continuous and dashed lines show the potential energy curves for BG in TG and sTG regime. In the trapless case, there is no potential depth for both TG and sTG regimes. In the TG regime, the condensate repeatedly shrinks and expand. In each expansion, atoms that are elastically scattered with highly energetic atoms cannot return to the stability region due to the absence of the external confinement potential. In this regime the condensate gradually decays (dispersive spreading of BG), hence the condensates expand, which is clearly illustrated in Figure 7 (left side). But, in sTG regime the interaction forces are attractive, so that the BG is stable up to finite time and it get collapses due to the high attraction. This point is illustrated in Figure 7 (right side). In Figure 7, we plot the numerical results for the trapless case, where the strength of the interaction $\xi_1(t)$ can be replaced by $\xi_f \left( a + b \sin(\omega t) \right)$ in equation (1). Here, the parameters $\xi_f$, $a$ and $b$ correspond to final, constant and coefficient of oscillatory part of the interactions, respectively.

From Figure 8, we show that the oscillation frequency increases the stability of the BG in sTG regime alone. But, in TG regime the BG expands to infinite due to both repulsive forces. In the sTG regime, the stability of the BG increases considerably for long time. This was due to the balance between the oscillatory force and the attractive force. This point is clearly presented in Figure 9.

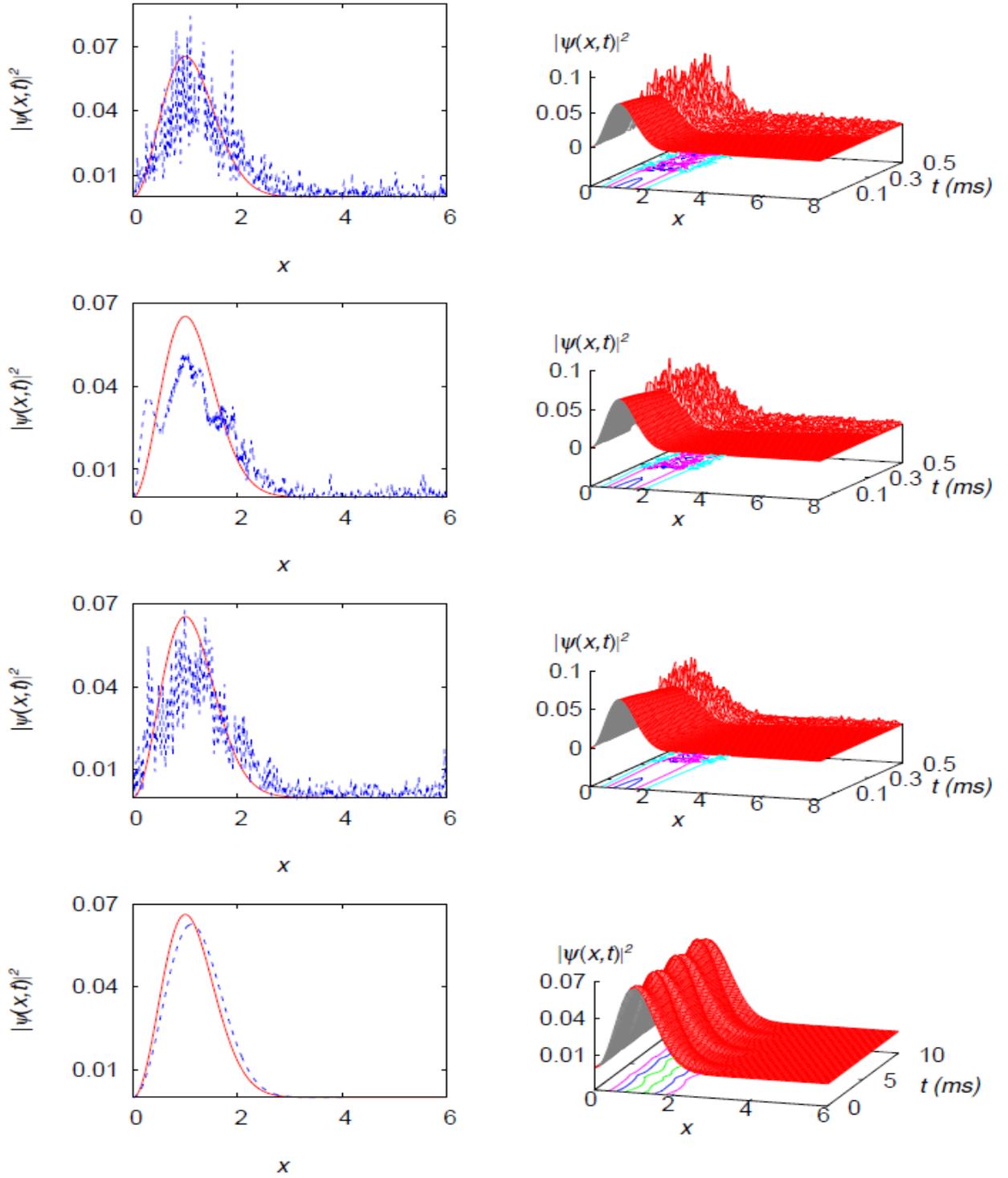

**Fig.9:** Stabilization of trapless BG in sTG regime in the presence of both constant and oscillatory parts of the interaction, $a = 1$, $b = 0.5$ and $\xi_f = -200$ in Equation (1). From upper to lower $\omega = 0\pi,\ 0.5\pi,\ 1\pi,\ 1.5\pi$

In Figure 10, the rms size of the radius $X_{rms}$ illustrates the stability of the trapless BG in sTG regime for different $\omega$ values, and the stability of the BG increases in sTG regime by tuning the oscillation part of the nonlinearity. It is clearly shown from Figures 9 and 10 that the analytical solution of Figure 8 is verified through numerical simulations. Hence, we conclude that the stability of the trapless BG in sTG regime can be increased by considering the oscillatory force in the system.

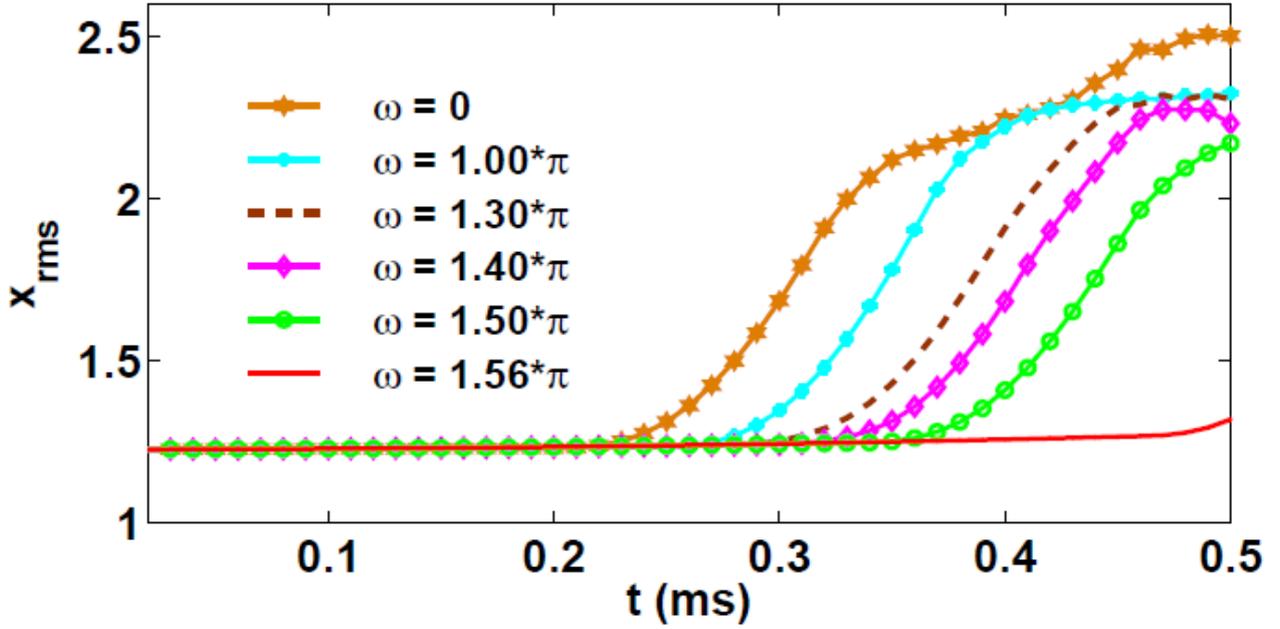

**Fig.10:** Plot of the *rms* radii versus *t(ms)* for fixed $\xi$ (=-200) and different $\omega$ values.

## IV- CONCLUSIONS

In summary, considering quintic Gross-Pitaevskii equation, we have theoretically analyzed the stability properties of Bose gas in both TG regimes with and without the external trapping potential. Once the model has been built, we have performed the variational approximation analysis and derived the equation of motion to investigate the stability of above two TG regimes in both cases. Abundant and interesting results were therefore obtained. For instance, based on the analytical framework, in the presence of external trapping potential, the Bose gas in the TG regime is found to be stable even for high interaction force or large number of atoms. But, in the case of sTG regime, the stability of the Bose gas decreases with increasing number of atoms in the system. We henceforth realize the stabilization of the Bose gas in sTG regime by appropriately selecting the strength of the external trap. However, when no external trap is applied, we have analyzed the stability properties of the Bose gas in both the TG regimes, and have then stabilize the Bose gas in sTG regime by considering an oscillatory force. Lastly, we have verified our analytical results through numerical simulations using the SSCN scheme for both afore-mentioned cases. The numerical results are deeply in agreement with those obtained using variational approximation method.

**ACKNOWLEDGEMENTS**

SS acknowledges the Foundation for Research Support of the State of Sao Paulo (FAPESP) [Contracts No. [2024/04174-8, 2020/02185-1, 2017/05660-0].

**COMPETING INTERESTS:**

The authors declare that they have no known competing financial and non-financial interests or personal relationships that could have appeared to influence the work reported in this paper.

**FUNDING**:

The authors declare that they have not received any funding from any organization to complete the work reported in this paper.

**AUTHOR CONTRIBUTIONS:**

R. Sasireka: Conceptualization, editing original draft, visualization, validation.

O.T. Lekeufack: Methodology, Software, writing, review, project administration, supervision.

A. Uthayakumar: Software, writing, visualization, review.

S. Sabari: Data curation, project administration, supervision, review.